\documentclass[final,5p,times,twocolumn]{elsarticle}

\usepackage{amssymb,amssymb,graphicx}
\usepackage{graphicx}
\usepackage{booktabs}

\journal{Journal of Magnetism and Magnetic Materials}

\newcommand{\p}{$\%$}
\newcommand{\pat}{{at.}\%~}

\newcommand{\tfn}{$\mathrm{Fe_{4}N}$}
\newcommand{\tcn}{$\mathrm{Co_{4}N}$}

\newcommand{\tmn}{$\mathrm{M_{4}N}$}
\newcommand{\Ts}{$\mathrm{T_{s}}$}

\begin{document}
\begin{frontmatter}
\title{Structure and Magnetization of \tcn~Thin Film}
\author {Nidhi Pandey$^1$, Mukul Gupta$^1$, Rachana Gupta$^2$, Parasmani Rajput$^3$ and Jochen Stahn$^4$}
\address{$^1$UGC-DAE Consortium for Scientific Research, University
Campus, Khandwa Road, Indore 452 001,India}
\address{$^2$Institute of Engineering and Technology DAVV, Khandwa Road, Indore 452 017,
India}
\address{$^3$Atomic and Molecular Physics Division, Bhabha Atomic Research
Centre, Mumbai 400 085, India }
\address{$^4$Laboratory for Neutron Scattering and Imaging, Paul
Scherrer Institut, CH-5232 Villigen PSI, Switzerland}
\address{$^*$ Corresponding author Email: dr.mukul.gupta@gmail.com/mgupta@csr.res.in\\Tel.:
+91 731 246 3913 \\Fax: +91 731 246 5437}

\begin{abstract}

In this work, we studied the local structure and the magnetization
of \tcn~thin films deposited by a reactive dc magnetron sputtering
process. The interstitial incorporation of N atoms in a fcc Co
lattice is expected to expand the structure and such expansion
yields interesting magnetic properties characterized by a larger
than Co magnetic moment and a very high value of spin polarization
ratio in \tcn. By optimizing the growth conditions, we prepared
\tcn~film having lattice parameter close to its theoretically
predicted value. The N concentration was measured using secondary
ion mass spectroscopy. Detailed magnetization measurements using
bulk magnetization method and polarized neutron reflectivity
confirm that the magnetic moment of Co in \tcn~is higher than that
of Co.

\end{abstract}


\end{frontmatter}

\section{Introduction}
\label{1} Transition metal nitrides (TMN) are an interesting class
of compounds as N incorporation in metals make them less corrosive
and results in interesting properties such as
superhardness~\cite{Veprek:Hard:99,PRL:Jhi:Vacancy:TMNs:01,Nature:Jhi:TMNs,Hao:PRL:superhard},
superconductivity~\cite{TMNs:superconductivity}, corrosion and
wear resistance~\cite{PRB:Steneteg:13:TMNs,Science:Sproul:hard}.
In particularly, tetra metal nitrides \tmn, generally formed only
for late 3d TM (e.g. Fe, Co, Ni)~\cite{JPCM_16_Markus} are
somewhat unique in the sense that they share a common fcc
structure. Due to a dominant metal-metal interactions in \tmn~they
possess metal like character (unlike metal
oxides)~\cite{JMMM10_Matar}. In a metal fcc cage, the
incorporation of N atoms at the interstitial positions, results in
an expansion of the unit cell, which in turn affects the magnetic
integrations \cite{PRB07_Matar, JMMM10_Imai, JAC14_Imai,
Coey.JMMM.1999, JMMM99_P_Mohn_Matar}. Generally incorporation of a
non-magnetic element in a magnetic one is expected to result in a
loss of magnetic moment ($M$) explained by the well-known
Slater-Pauling curve ~\cite{JMMM10_Matar}. However, theoretical
calculations predict $M$ to be larger for \tfn~and \tcn~compounds.
This is an interesting proposition as with larger than metal $M$,
corrosion resistance and a metal like character can make
\tmn~compounds an alternative their pure metal counterparts. In
addition, some recent theoretical predicted a very spin
polarization ratio (SPR) for \tmn~compounds\cite{PRB07_Matar,
JMMM10_Imai}. Among them \tcn~is predicted to have SPR $\sim$
90\p~\cite{JMMM10_Imai} which is among the highest values for any
compound.

However experimental results obtained so far for \tcn~compounds
does not seems to be as exciting as theoretical calculations. This
can be understood from the fact that the value of $M$ found in
most of the experimental works so far is perpetually less than
than that of pure Co. A closer look at the \tcn~thin films reveal
that the values of lattice parameter (LP) obtained in most of the
experimental works is typically 3.54\,{\AA}~\cite{Silva2015,
2011_Co4N_K_Ito, JAP14_Ito, TSF14_Silva, JMS87_Oda,
Wang:CoN:TSF:09, MSEB08_Jia}. This is more close to theoretical
values LP for fcc Co at 3.54\,{\AA} than that of \tcn~at
3.72\,{\AA}~\cite{PRB07_Matar, JMMM10_Imai, Silva2015}. Such a
discrepancy in the experimental and theoretical values of LP for
\tcn~requires attention. While investigating the recipes adopted
for formation of \tcn~thin films, we found that \tcn~films were
often deposited at substrate temperatures (\Ts) similar to those
used for \tfn~thin films. This is convenient approach due to
similarity between the \tfn~and \tcn~and the absence of a phase
diagram for the system Co-N, intuitively makes one to follow paths
adopted for preparation of \tfn.

However, the energetics of nitride formation for \tfn and
\tcn~immediately indicates about the complexity for the later.
Theoretical values of enthalpy of formation
($\Delta$$H^{\circ}_{f}$) for \tfn~is about
-12\,kJ~mol$^{-1}$~\cite{Tessier_SSS00}, whereas those for
\tcn~are slightly above or below 0 for hcp Co or fcc Co
~\cite{JMMM10_Imai}. This also implies that at a higher \Ts,
\tcn~system will be less stable as compared to \tfn. In a recent
work, we studied the phase formation process in the Co-N system at
\Ts~= 300\,K~\cite{CoN_AIP_Adv2015} and
523\,K~\cite{JAC16_NPandey}. We found that at \Ts~= 523\,K, N
incorporation in the Co-N system is minimal and the phases formed
are similar to a fcc Co having LP$\sim$3.52{\AA}. On the other
hand when \Ts is lowered to 300\,K Co-N depicts a similar type of
phase formation sequence as found for the Fe-N
system~\cite{Gupta:PRB05, JAC15_ATayal}. By optimizing the
deposition conditions, \tcn~film having LP as high as 3.68\,{\AA}
can be deposited and the value of $M$ also supersedes that of Co.
Since the \tcn~films deposited without any intentional heating are
expected to have a large fraction of disorder, estimation of LP
with x-ray diffraction alone may not be decisive, therefore in the
present work, we investigate the local structure and $M$ of the
\tcn~thin film deposited at \Ts~= 300\,K using x-ray absorption
based techniques. By doing measurements at Co K and L-edges and at
N K-edge, we get valuable information about the local structure.
In addition, by doing polarizer neutron reflectivity (PNR)
measurements at low temperatures ($\sim$20\,K) we determined the
value of $M$ for \tcn~thin film. Obtained results are presented
and discussed in this work.

\section{Experimental}
\label{2} We deposited \tcn~thin film with a nominal thickness of
120\,nm at\Ts~= 300\,K (without intentional heating) using a
reactive direct current magnetron sputtering (dcMS) system
(Orion-8, AJA Int. Inc.). A one inch diameter and 0.5\,mm thick
pure Co (purity 99.95\p) target was sputtered using a gas mixture
of Ar and N$_2$ (both 99.9995\% pure) gases. With a base pressure
of 1$\times$10$^{-7}$\,Torr, the pressure during deposition was
about 3\,mTorr. More details about deposition process can be found
in ~\cite{CoN_AIP_Adv2015}. Along with the \tcn~sample, we also
deposited a pure Co thin film under identical conditions as a
reference.

To investigate the local and electronic structure, x-ray
absorption near edge spectroscopy (XANES) and extended x-ray
absorption fine structure measurements (EXAFS) were performed in
the total electron yield (TEY) and florescence mode at
BL-01~\cite{XAS_beamline} and BL-09 beamlines, respectively at the
Indus-2 synchrotron radiation source at RRCAT, Indore. The
measurements in TEY mode at BL-01 were carried out in a UHV
chamber with a base pressure of (2$\times$10$^{-10}$\,Torr). To
avoid surface contaminations, samples were cleaned $in-situ$ using
a Ar$^+$ source kept incident at an angle of 45 $^{\circ}$. The
measurement in florescence mode at BL-09 were carried out at
ambient conditions.

The composition of \tcn~thin film was measured using secondary ion
mass spectroscopy (SIMS) depth profiling using a Hiden Analytical
SIMS workstation. An oxygen ion beam of energy 4\,keV and 200\,nA
was used as a primary source and the sputtered species were
detected using a quadrupole mass analyzer. The SIMS depth profiles
were compared with a references sample as described in
~\cite{CoN_AIP_Adv2015}. The magnetization measurements were
carried out using a Quantum Design SQUID-VSM (S-VSM) magnetometer
at room temperature. We did PNR measurements at AMOR
reflectometer~\cite{Gupta_PramJP04} in the time of flight mode at
SINQ-PSI Switzerland. To saturate the sample magnetically, a
magnetic field of 0.5\,T was applied during the PNR measurements.
The measurements at low temperature were carried out using a close
cycle refrigerator installed inside the electro magnet.

\begin{figure} \center
\vspace{-5mm}
\includegraphics [width=60mm,height=45mm] {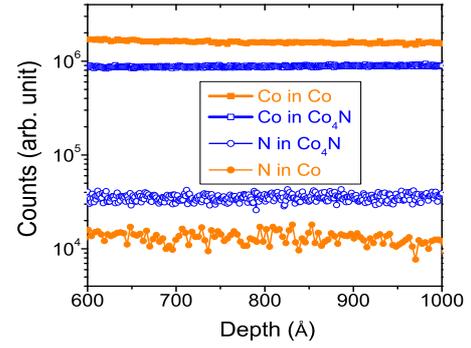}
\caption{\label{fig:SIMS} (Color online) SIMS depth profiles of Co
and \tcn~thin films samples.} \vspace{-2mm}
\end{figure}

\section{Results and Discussion}
\label{3}

To quantify the N \pat in our samples, we did SIMS depth profiling
measurements. The SIMS depth profiles for the \tcn~sample are
shown in fig.~\ref{fig:SIMS} along with a Co reference sample. The
depth profiles clearly reveal that N concentration in the
\tcn~sample is more and the Co concentration is less as compared
to Co sample. Following a procedure described in
ref.~\cite{CoN_AIP_Adv2015} and measuring a reference sample with
known concentration, we found that N \pat~comes out to be
$\sim$18($\pm$2)\pat~indicating formation of \tcn~phase.

\begin{figure}\center
\vspace{-1mm}
\includegraphics [width=85mm,height=37mm] {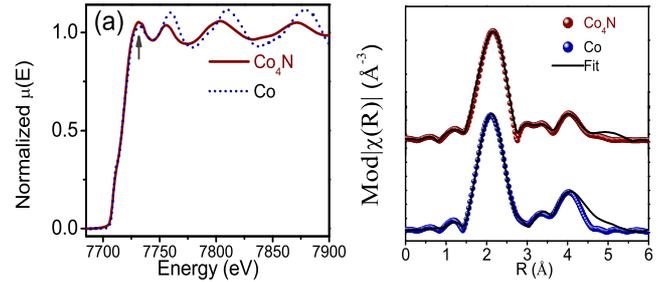}
\caption{\label{fig:XAFS} (Color online) Normalized Co K-edge
XANES spectra of \tcn~and Co foil samples (a). Fitted k$^2$
weighted spectra for \tcn~and a Co samples (b).} \vspace{-3mm}
\end{figure}

\begin{figure}\center
\vspace{-0.1mm}
\includegraphics [width=75mm,height=50mm] {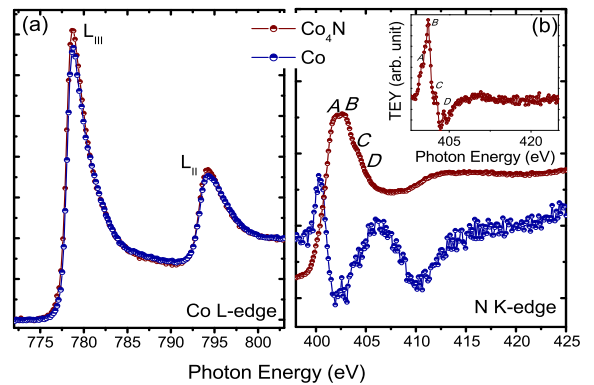}
\caption{\label{fig:XAS} (Color online) XAS spectra of \tcn~and Co
thin film at Co-L-edge (a) at N K-edge (b). The inset in (b) shows
the derivative of N K-edge in \tcn.} \vspace{-3mm}
\end{figure}

\begin{figure*}\center
\vspace{-5mm}
\includegraphics [width=120mm,height=100mm] {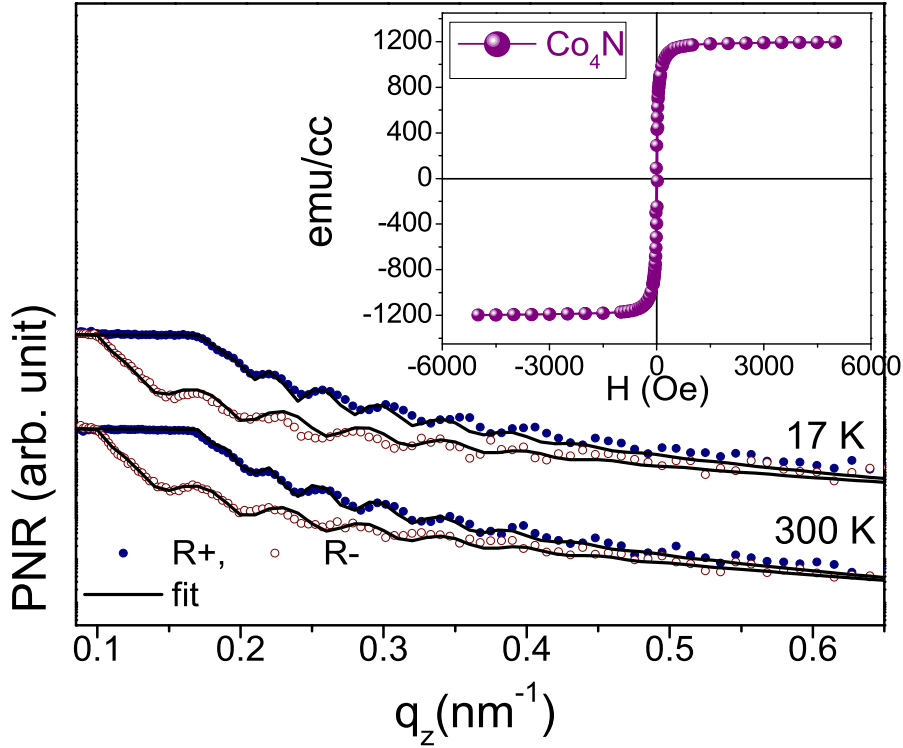}
\caption{\label{fig:PNR} (Color online) (a). PNR patterns of \tcn~
thin film taken at room temperature and 17\,K. Inset showing the
M-H curve at room temperature of \tcn.} \vspace{-5mm}
\end{figure*}

X-ray diffraction measurements (not shown) carried out in our
samples resulted in similar patterns as observed in a earlier
work~\cite{CoN_AIP_Adv2015}. Since samples are deposited without
any intentional heating, the long range structure is expected to
somewhat disordered and more reliable information about the
structure can be obtained from a local probe like x-ray absorption
spectroscopy. To get complementary information, we did XANES
measurements at Co K and L-edges and at the N K-edge. In addition,
EXAFS measurements were also carried out at Co K-edge.
Fig.~\ref{fig:XAFS}(a) and (b) shows the normalized XANES and
EXAFS patterns for \tcn~sample together with a Co metal as a
reference. It can be clearly seen that the intensity in the white
line region is higher in the \tcn~sample(marked by an arrow in
~\ref{fig:XAFS}(a)), which essentially arises due to hybridization
between face centered Co 3d and N 2p bands through $\sigma$$^*$
anti-bonding and an enhancement in density of states in \tcn~as
compared to Co. Such effects are even more pronounced at Co
L-edges as shown in fig.~\ref{fig:XAS}(a). Here presence of N in
\tcn~thin film can also be confirmed by comparing the N K-edge
spectrum of \tcn with that of a Co reference sample as shown in
fig.~\ref{fig:XAS}(b). The features in N K-edge spectra are
labelled as $A$, $B$ and $C$. Here the feature $A$ is attributed
to a dipole transition from N 1s to hybridized states of face
centered Co 3d and N 2p orbitals through $\pi$$^*$ anti-bonding.
Features $B$ and $C$ represents the dipole transition from N 1s to
hybridized states of face centered Co 3d and N 2p orbitals through
$\sigma$$^*$ anti-bonding~\cite{APL_11_K_Ito_Co4N,
JAP15_KIto_Fe4N}. These results obtained from XAS measurements
provide a strong evidence for the \tcn~structure.

Further information about the LP can be obtained from EXAFS
measurements. Fig.~\ref{fig:XAFS}(b) shows the modulli of k$^2$
weighted EXAFS data for \tcn~and with a Co reference. For
\tcn~structure we consider space group $Pm\bar{3}m$ and LP =
3.74\,{\AA}. Among other parameters the fitting of EXAFS data also
yields the distance between corner Co atoms which is equal to the
LP. We found that the LP comes out to be 3.92~\,{\AA} with
($\sigma$$^2$ = 0.0145~\,{\AA}$^2$); these values for the Co
reference sample are 3.47~\,{\AA} with ($\sigma$$^2$ =
0.0093~\,{\AA}$^2$). The value of LP obtained in our case is
considerably larger than that of Silva $et. al.$ ~\cite{Silva2015}
in a recent work for their \tcn~thin film. They found LP =
3.53~\,{\AA} with ($\sigma$$^2$ = 0.005~\,{\AA}$^2$). Considering
even large inaccuracy of EXAFS technique in determination of LP,
our values are considerably larger and clearly indicate an
expansion in the Co lattice by incorporation of N. It may be noted
that Silva $et. al.$ ~\cite{Silva2015} deposited their samples at
a \Ts~= 523\,K and at this temperature it may happen that most of
N atoms diffuse out of the system leaving behind a dominant fcc Co
as found in our recent work~\cite{JAC16_NPandey}.

It may be noted that in most of experimental work carried out in
the \tcn~thin films always a high \Ts~was used and the measured
values of LP were closer to fcc Co rather than that of \tcn. Since
LP and the $M$ are correlated and an expansion in the LP due to
interstitial incorporation of N atoms is responsible for enhanced
$M$~\cite{JMMM10_Imai}, the values of $M$ obtained in most of the
works were closer to pure Co. To measure $M$ in our samples we did
bulk magnetization and PNR measurements. Inset of
fig.~\ref{fig:PNR} shows M-H loop measured using S-VSM at room
temperature for a \tcn~thin film. We find that $M$ = 1195\,emu/cc
which approximately corresponds 1.6$\mu_\mathrm{B}$ per Co atom,
which is slightly lower than the theoretically predicted value of
$M$ for \tcn. Exact determination of $M$ from bulk magnetization
measurements in a thin film sample require precise values of
sample volume and density. While the former can be measured with a
great accuracy, estimation of later is not easy due to relatively
large fraction of defects etc. in thin films. In addition
diamagnetism of substrate always interferes with sample
magnetization.

It is well-known that in a PNR measurement, absolute value of $M$
can be measured as in PNR technique, measurement of $M$ is not
influenced by sample volume and substrate magnetism. On the other
hand density of the film is inherently measured in PNR. In this
context it is surprising to note that PNR has not been used to
measure $M$ in \tcn~thin films. In most of the works available so
far, $M$ was measured using bulk magnetization methods and there
seems to be a large variations in estimation of $M$ for \tcn~thin
films ranging from 1.3 to 1.6$\mu_\mathrm{B}$/atom
~\cite{JAP14_Ito,JMS87_Oda,Wang:CoN:TSF:09,MSEB08_Jia,APL_11_K_Ito_Co4N}.

We performed PNR measurement at room temperature and at 17\,K (to
minimize thermal fluctuations) under an applied magnetic field of
0.5\,T, which is sufficient to saturate sample magnetically (see
inset of~\ref{fig:PNR}). The spin-up down reflectivities clearly
show a separation typically expected for a ferromagnetic sample.
The fitting of PNR patterns was carried out using SimulReflec
programme~\cite{SimulReflec} and the obtained values of $M$ are
1.73($\pm$0.05)\,$\mu_\mathrm{B}$/atom at 300\,K and
1.75($\pm$0.05)$\mu_\mathrm{B}$/atom at 17\,K. This clearly shows
that $M$ is larger than Co in \tcn~thin film. Though such
enhancement in $M$ was theoretically predicted, it has been
unambiguously demonstrated in this work.

\section{Conclusion}
\label{4}

In conclusion, by measuring the local structure and a we found
that LP of \tcn~thin films far exceeds that of previous works and
is more closer to its theoretical value. For \tcn~thin films an
enhancement in $M$ was expected. By by doing precise magnetization
measurements using PNR we found enhancement in the Co magnetic
moment. In addition, our SIMS depth profile measurements cleanly
reveal the formation of \tcn~phase and N K-edge measurements
further confirms the formation of \tcn~phase.

\section{Acknowledgments}
A part of this work was performed at AMOR, Swiss Spallation
Neutron Source, Paul Scherrer Institute, Villigen, Switzerland. We
acknowledge Department of Science and Technology, New Delhi for
providing financial support to carry out PNR experiments. We are
thankful to Layanata Behera for the help provided in sample
preparation and various measurements. We acknowledge Ram Janay
Choudhary and Malvika Tripathi for S-VSN measurements. Thanks are
due to D. M. Phase and Rakesh Sah for support in XAS beamline. We
are thankful to V. Ganesan and A. K. Sinha for support and
encouragements.\\

%\section{References}
%\bibliography{TMN}

\end{document}